\documentclass[12pt]{article}
\usepackage{amssymb} 

\newcommand{\R}{{\mathbb R}}
\newcommand{\C}{{\mathbb C}}
\renewcommand{\H}{{\mathbb H}}
\renewcommand{\ltimes}{{\kern3pt\hbox{\vrule width 0.4pt height
5.30pt depth .0pt}\kern-1.76pt\times\kern1pt}}

\begin{document}


\begin{titlepage}
\begin{flushright}
MCTP-04-08
\\hep-th/0402141
\end{flushright}

\vspace{15pt}

\begin{center}
{\large\bf Generalized Holonomy of Supergravities with $8$ Real
Supercharges\footnote{Research supported in part by DOE Grant
DE-FG02-95ER40899.}}

\vspace{15pt}

{A.~Batrachenko\footnote{abat@umich.edu} and W. ~Y.
~Wen\footnote{wenw@umich.edu}}

\vspace{7pt}

{\it Michigan Center for Theoretical Physics\\
Randall Laboratory, Department of Physics, University of Michigan\\
Ann Arbor, MI 48109--1120, USA}
\end{center}

\begin{abstract}

We show that the generalized holonomy groups of ungauged
supergravity theories with $8$ real supercharges must be contained in
$SL(2-\nu,\H)\ltimes{\nu\H^{2-\nu}}\subseteq SL(2,\H)$, where $SL(2,\H)$
is the generalized structure group.  Here $n=4\nu$ is the number of
preserved supersymmetries, so the allowed values are limited to $n=0,4,8$.
In particular, solutions of ungauged supergravities in
four, five and six dimensions are examined and found to explicitly follow
this pattern.  We also argue that the $G$-structure has to be a
subgroup of this generalized holonomy group, which may provide a
possible classification for supergravity vacua with respect to the
number of supercharges.
\end{abstract}

\end{titlepage}

\section{Introduction}
For a general solution of supergravity theory in dimension $D$,
the number of supersymmetries preserved by this background depends
on the number of covariantly constant spinors,
\begin{equation}
\label{covar}
{\cal D}_M\varepsilon=0,
\end{equation}
Generically, ${\cal D}_M$ includes the covariant derivative $D_M$
with Levi-Civita connection $\omega$ and contribution from flux $F$.
For the simpler vacua where flux vanishes, one
obtains the integrability condition,
\begin{equation}
\label{intgra}
[D_M,D_N]\varepsilon=\frac{1}{4}{R_{MN}}^{AB}\Gamma_{AB}\varepsilon=0
\end{equation}
where ${R_{MN}}^{AB}$ is Riemann tensor and $\Gamma_{AB}$ can be
viewed as generators of structure group ${\cal G}=SO(1,D-1)$. The
holonomy group ${\cal H}$ associated with $\omega$ is a subgroup
of ${\cal G}$. The number of preserved supersymmetries is the
number of singlets appearing in the decomposition of the spinor
representation of $SO(1,D-1)$ under ${\cal H}$. The holonomy
groups in Lorentzian signature has been studied \cite{Bryant} and
in Euclidean signature they have been completely classified
\cite{Berger}.

For generic backgrounds, however, one has to consider the
contribution from flux; hence the classification above is no
longer valid. A statement that an enlarged structure group
$\tilde{\cal G}$ and a generalized holonomy group $\tilde{\cal H}$
exist to play similar roles as in the pure (pseudo) Riemannian
background has been verified for M-theory vacua in eleven
dimensions
\cite{Duffstelle,Duff,Duff:01,Hull,Papadopoulos,Batrachenko} and
type IIB theory in ten dimensions \cite{Papadopoulos:01}. It would
be interesting to see if this approach could be applied to
supergravity theories in lower dimensions and if the holonomy
groups of theories in different dimensions are related. In this
paper, we argue that a single generalized structure group, with
corresponding generalized holonomy subgroups is responsible for
classification of supergravity vacua with $8$ real supercharges.
In particular, all solutions of ungauged supergravity in four,
five, and six dimensions are investigated to support our
proposition. While the $G$-structure is useful to construct
solutions, the generalized holonomy approach may provide a better
classification of supersymmetry vacua with regard to preserved
supercharges, at least for the vacua examined in this paper.

This paper is organized as follows. In section \ref{gh8susy}, we
show that the generalized structure group for supergravities with $8$
supercharges in $D=4,5,6$ is $SL(2,\H)$. Then we argue that the
generalized holonomy is at least contained in $SL(1,\H)\ltimes{\H}$ for
all $1/2$-BPS
solutions. In section \ref{gh5d}, as an example, we examine the
generalized holonomy of minimal five-dimensional supergravity
solutions of both timelike and null cases. Finally, investigation
of four- and six-dimensional supergravity vacua provides more
evidence and this discussion is carried out in section \ref{discuss}.

\section{Generalized holonomy of supergravity solutions with $8$ real
supercharges}
\label{gh8susy}
Here we are interested in supergravity theories
with $8$ real supercharges, specifically, $\mathcal N=(1,0)$ in $D=6$,
$\mathcal N=2$ in $D=5$, $\mathcal N=2$ in $D=4$  dimensions.  They are
closely
related in the sense that the last two theories can be obtained from
the first one through Kaluza-Klein reduction with consistent
truncation \cite{Ortin,Ortin:01}. The (pseudo) Riemannian structure group is generated by
gamma matrices with two indices, denoted as $\Gamma^{(2)}$, as
shown in (\ref{intgra}), which is $SO(1,D-1)$ for each $D=4,5,6$.
In the more general case of non-zero flux $F$, one considers the
generalized structure group using the Killing spinor equation with
flux $F$ turned on, which can be written schematically as
\begin{equation}
\label{kseq}
{\cal D}_M \varepsilon= [D_M +
(\Gamma^{(p+1)}F_{(p)})_M+(\Gamma^{(p-1)}F_{(p)})_M]\varepsilon,
\end{equation}
where $F_{(p)}$ is a two-form in $D=4,5$ and a three-form in
$D=6$.  Hence, in (\ref{kseq}) we find the combinations $\Gamma^{(1)}$,
$\Gamma^{(2)}$, $\Gamma^{(3)}$ for $D=4,5$ and $\Gamma^{(2)}$,
$\Gamma^{(4)}$ for $D=6$.

One can compute the independent gamma matrix combinations for each
specific dimension. For $D=4$, choosing
the chirality projector $\Gamma^5\varepsilon=\varepsilon$, the
integrability condition gives one more generator $\Gamma^{(4)}$,
and as a result $15$ independent generators in total. For $D=5$,
the fact $\Gamma^{(2)}$ is dual to $\Gamma^{(3)}$ gives us exactly
$15$ generators.  As to $D=6$ case, by choosing the chirality
projector $\Gamma^7\varepsilon=\varepsilon$, one only obtains $15$
relevant generators for the $\mathcal N=(1,0)$ theory.  The $15$ in
each of the three cases
forms a real Clifford group isomorphic to $SL(2,\H)$.  Hence
this is the generalized structure group with regard to $8$
supercharges, at least for the three theories just mentioned.

To find the generalized holonomy group for vacua preserving a subset
$0\le n\le8$ of supersymmetries, one considers the subgroup of
$SL(2,\H)$ that stabilizes a quaternion-valued spinor of the form
$\pmatrix{s^1&s^2}$, where each $s^i$ has four real components.
Consider, for example,
\begin{equation}
\label{stable}
\pmatrix{A&B\cr{C}&{D}}\pmatrix{s^1\cr s^2},
\end{equation}
where $A,B,C,D\in \H$.  We see that for arbitrary $A,B,C,D$ there
is no preserved spinor ($n=0$). On the other hand, the trivial
identity ($A=D=1, B=C=0$) will preserve both $s^1$ and $s^2$
($n=8$). The only nontrivial choice is $A=1, C=0$ and $D\in
SL(1,\H)$, which preserves half the supercharges ($n=4$), say
$s^1$ in this choice.  This demonstrates that the number of
preserved supersymmetries is restricted to the values $n=0,4,8$,
corresponding to solutions with no supersymmetry, half and maximal
supersymmetry, respectively. This observation agrees with
\cite{Gauntlett} in $D=5$ ungauged supergravity case. In
conclusion, for a solution to preserve exactly $n$ supersymmetries
the generalized holonomy ${\cal H}$ has to satisfy the condition
$SL(\frac{4-n}{4},\H)\ltimes
\frac{n+4}{4}{\H}^{\frac{4-n}{4}}\subset {\cal H}\subseteq
SL(\frac{8-n}{4},\H)\ltimes \frac{n}{4}{\H}^{\frac{8-n}{4}}$.

To complete our discussion here, we note that the $SL(2,\H)$ classification
only holds for the ungauged supergravities.  For gauged supergravities,
a gauge term such as $igA_\mu$, with coupling $g$, has to be included in
the generalized covariant derivative (\ref{kseq}). This modifies the
generalized structure group to the complex Clifford group $GL(4,\C)$ with
$32$ generators.  However, as we have seen above, it is restricted to the
real Clifford group $SL(2,\H)$ in the ungauged theory.

\begin{table}
\begin{center}
\begin{tabular}{cccc}
Spinor&Maximal supersymmetry&Maximal generalized holonomy\\
\hline
Real&$32$&$SL(32-n,\R)\ltimes n{\R}^{32-n}$\\
Complex&$8$&$GL(\frac{8-n}{2},\C)\ltimes \frac{n}{2}{\C}^{\frac{8-n}{2}}$\\
Quaternion&$8$&$SL(\frac{8-n}{4},\H)\ltimes \frac{n}{4}{\H}^{\frac{8-n}{4}}$\\
\end{tabular}
\end{center}
\caption{Generalized holonomies of supergravities with $8$ and $32$
real supercharges.  Here $n$ is the number of preserved supersymmetries.}
\label{gh}
\end{table}

Table \ref{gh} summarizes our findings together with another
fundamental result for the generalized holonomy of M-theory
\cite{Hull,Papadopoulos}. The first category in table \ref{gh}
is useful for discussion of supergravity theories with $32$ real
supercharges or fewer \cite{Batrachenko} where the spinors take
values in $\R$. The second and third categories are useful for
discussion of supergravity with $8$ real supercharges or less where
the spinors may take values in either $\C$ or $\H$. To be specific,
we may classify the generalized holonomy of $D=5$ gauged
supergravity solutions according to the second category.  We will
return to this problem later in section \ref{discuss}.

In the following section, as an example, we will explicitly compute
the generalized holonomy group for $D=5$ minimal supergravity to support
the classification scheme of table~\ref{gh}.

\section{Generalized holonomy of $D=5$ minimal supergravity}
\label{gh5d}


\subsection{$D=5$ minimal supergravity}
\label{sugra}

All supersymmetric solutions of minimal supergravity in five
dimensions have been classified using the idea of $G$-structure
\cite{Gauntlett}.  Here we follow their convention and give a
brief review as follows. The bosonic action for minimal
supergravity in five dimensions is
\begin{equation}
S = \frac{1}{4\pi G} \int{-\frac{1}{4} R\star 1 - \frac{1}{2}F\wedge\star F
-\frac{2}{3\sqrt{3}}F\wedge F\wedge A},
\end{equation}
If a solution to the equation of motion derived from the above
action is supersymmetric, it admits a covariantly constant spinor
satisfying
\begin{equation}
{\cal
D}_M\varepsilon^a=[D_M+\frac{1}{4\sqrt{3}}({\Gamma_M}^{NP}-4\delta_M^N\Gamma^P)F_{NP}]\varepsilon^a=0,
\end{equation}
where $\varepsilon^a$ is a symplectic Majorana Killing spinor with
$8$ real components. Out of this Killing spinor, one can construct a real
scalar $f$, a real $1$-form $V$ and three complex $2$-forms $\Phi$ as
\begin{eqnarray}
\label{kforms}
f\epsilon^{ab}&=&\bar{\varepsilon}^a\varepsilon^b, \nonumber\\
V_M\epsilon^{ab}&=&\bar{\varepsilon}^a\Gamma_M\varepsilon^b,
\nonumber\\
\Phi_{MN}^{ab}&=&\bar{\varepsilon}^a\Gamma_{MN}\varepsilon^b,
\end{eqnarray}
There are also algebraic and differential constrains for these
objects to satisfy, which we will not detail here. It turns out
that $V_M$ is a Killing form (vector) satisfying
\begin{equation}
V_MV^M=f^2,
\end{equation}
and the solutions can be classified according to whether $V_M$ is
timelike ($f\neq 0$) or null ($f=0$).  In the following, we
examine both cases.

\subsection{The timelike solution}
\label{time} By choosing the timelike killing vector
$V=\partial/\partial t$, the metric, in general, can be written
locally as \cite{Gauntlett}
\begin{equation}
\label{5dtime}
ds^2=f^2(dt^2+\omega)^2-f^{-1}h_{ij}dx^idx^j,
\end{equation}
where $h_{ij}$, $i,j=1..4$ is the metric of a hyper-K\"{a}hler base manifold
$\cal B$
together with a globally defined $f$ and locally defined 1-form connection
$\omega$ on $\cal B$.
The two form $d\omega$ can then be split into self-dual
and anti-self-dual parts with respect to $h_{ij}$:
\begin{equation}
fd\omega=G^+ + G^-
\end{equation}
It is convenient to introduce a local (flat) frame such that
\begin{eqnarray}
e^0=f(dt+\omega), \nonumber\\
e^ie^j=f^{-1}h_{ij}dx^idx^j
\end{eqnarray}
Then the two-form flux can be written as
\begin{equation}
F=\frac{\sqrt{3}}{2}de^0-\frac{1}{\sqrt{3}}G^+.
\end{equation}
The Bianchi condition and equation of motion give the constrains:
\begin{eqnarray}
dG^+ &=& 0,  \nonumber\\
\Delta f^{-1} &=& \frac{2}{9}{G^+}_{ij}{G^+}^{ij},
\end{eqnarray}
where $\Delta$ is the Laplacian with respect to $\cal B$.

Using the metric given by (\ref{5dtime}) and the constrains above, we
can obtain the Killing spinor equations:
\begin{eqnarray}
{\cal D}_t \varepsilon &=& [\partial_t + f^{1/2}\partial_i f \gamma^iP_-
-\frac{1}{6}f^2G^+\cdot\gamma P_-]\varepsilon,  \\
{\cal D}_i \varepsilon &=& [\partial_i -
\frac{1}{2}f^{-1}\partial_i f +
\omega_i(f^{1/2}\partial_jf\gamma^j-\frac{1}{6}f^2G^+\cdot\gamma)P_-
\nonumber\\
&&-\frac{1}{2}f^{-1}\partial_jf({\gamma_i}^j-2\delta_i^j)P_-
-\frac{1}{3}f^{1/2}(G^++3G^-)_{ij}\gamma^jP_-]\varepsilon,
\end{eqnarray}
where the gamma matrices are in the local frame and obey
\begin{eqnarray}
\label{gamma}
\{\gamma_A,\gamma_B\}=2\eta_{AB},\nonumber\\
\gamma_{ABCDE}=\epsilon_{ABCDE}.
\end{eqnarray}
Here we choose $\eta_{00}=1$, $\eta_{ij}=-\delta_{ij}$, and
$\epsilon_{01234}=1$.  $P_\pm=\frac{1}{2}(1\pm\gamma^0)$ is the
$\frac{1}{2}$-BPS projection for the timelike background.
To obtain the generalized holonomy, we examine the commutator of covariant
derivatives. Defining
\begin{equation}
\label{commu}
{\cal M}_{MN}\varepsilon=[{\cal D}_M,{\cal D}_N]\varepsilon,
\end{equation}
we find
\begin{eqnarray}
{\cal M}_{it}\varepsilon&=&{\cal A}_{ij}\gamma^jP_-
+ {\cal B}_{ijk}\gamma^{jk}P_-\varepsilon, \nonumber\\
{\cal M}_{ij}\varepsilon&=&{\cal C}_{ijm}\gamma^mP_-
+ {\cal D}_{ijmn}\gamma^{mn}P_-\varepsilon,
\end{eqnarray}
where ${\cal A}, {\cal B}, {\cal C}, {\cal D}$ only depend on
functions $f$ and $G$.  For example, $\mathcal A$ and $\mathcal B$ are
given by
\begin{eqnarray}
{\cal A} &=& f^{1/2}\partial_i\partial_jf -
\frac{1}{2}f^{-1/2}\delta_{ij}(\partial f)^2 +
\frac{2}{9}f^{5/2}(G^++3G^-)_{ik}{{G^+}_j}^k  \nonumber\\
{\cal B} &=& -\frac{1}{6}f^2\partial_i {G^+}_{jk}
- \frac{1}{3}[f\partial_{(i}f{G^+}_{j)k}-\delta_{ij}\partial_lf{G^+}_{lk}].
\end{eqnarray}
Since the details are not important for finding the generalized holonomy, we
do not provide the more complicated expressions for $\mathcal C$ and
$\mathcal D$.

We see that the only combination of gamma matrices showing up in
${\cal M}_{MN}$ are given by $\gamma^{i}P_-$ and $\gamma^{ij}P_-$.
Defining two sets of generators
\begin{eqnarray}
T^{ij}=-\frac{i}{2}P_-\gamma^{ij}P_-, \nonumber\\
K^{i}=P_+\gamma^iP_-,
\end{eqnarray}
and observing that
$\frac{1}{2}\epsilon_{ijkl}\gamma^{kl}=-\gamma_{ij}\gamma^{0}$, we
find that the only independent generators for $T^{ij}$ are
$T^{12}$, $T^{23}$ and $T^{31}$.  They generate the $SU(2)_-$
algebra, where the $-$ refers to the sign of the $P_-$ projection.
The other generators $K^i$ obviously commute among themselves due
to the projection identity $P_+P_-=0$.  If we choose $T^{31}$ as the
Cartan generator for
$SU(2)$, we may see that $\{K^1,K^3\}$ and $\{K^2,K^4\}$ form two
doublets.  Therefore the generalized holonomy for the timelike
solutions is
\begin{equation}
{\cal H}_{timelike}=SU(2)_-\ltimes 2\R^{2}.
\end{equation}
%

\subsection{The null solution}
\label{null}
When the Killing vector V defined in (\ref{kforms}) is null it is
possible to choose coordinates ($u$,$v$,$y_i$), $i$=1,2,3, such
that $V$ is tangent to geodesics in the surface of constant $u$
with affine parameter $v$, i.e. $V=\frac{\partial}{\partial v}$.
In this set of coordinates the most general metric obeying the algebraic
identities that relate the components of the forms (\ref{kforms})
is \cite{Gauntlett}
\begin{equation}
\label{nullmeric}
ds^2=H^{-1}(u,{\bf x})({\cal F}(u,{\bf x})du^2+2dudv)-H^2(u,{\bf x})(d{\bf
x}+{\bf a}(u,{\bf x}) du)^2.
\end{equation}
The field strength is given by
\begin{eqnarray}
&F_{ui}=-\frac{H^{-2}}{4 \sqrt 3} \epsilon_{ijk} \partial_j
\left(H^3 a_k \right),\\
&F_{ij}=-\frac{\sqrt 3}{4} \epsilon_{ijk} \partial_k H.
\end{eqnarray}

The generalized covariant derivative in this background takes the form
\begin{eqnarray}
\label{nullDs} {\cal D}_v\varepsilon&=&\left(\partial_v+\frac 12
H^{-2}\partial_aH\gamma^{a+}P_-\right)\varepsilon, \\
{\cal D}_u\varepsilon&=& \left[\left( \partial_u - a^i \partial_i
\right) - \frac 14 \partial_i \left({\cal F}H^{-1} \right)
\gamma^{i+}P_- + \frac 13
\epsilon_{ijk} \partial_j a_k \gamma^i P_-\right] \varepsilon,\\
{\cal D}_i \varepsilon &=&
\left[\partial_i-H^{-1}\epsilon_{ijk}\partial_jH\gamma^kP_--\frac16
H^2\epsilon_{ijk}\partial_j a_k\gamma^+ P_-\right.\nonumber\\
+&&\left.\left(\frac 13 H^2\partial_j a_i+\frac 16 H^2
\partial_i a_j-\frac 12\delta_{ij} H (\partial_uH-a_k\partial_kH)\right)
\gamma^{j+}P_- \right]\varepsilon.
\end{eqnarray}
Here the $\frac 12$-BPS projectors are defined as $P_-=\frac 12
\gamma^- \gamma^+$, $P_+=\frac 12 \gamma^+ \gamma^-$. The gamma
matrices are defined in (\ref{gamma}) where $\eta_{+-}=\eta_{-+}=1$,
$\eta_{ij}=-\delta_{ij}$ and $\epsilon_{+-123}=1$.

The action of the holonomy group on an
arbitrary spinor $\varepsilon$ is represented by the commutator of
the generalized covariant derivatives as defined in (\ref{commu})
\begin{eqnarray}
 {\cal M}_{vi}\varepsilon &=& {\cal A}_{ij}\gamma^{j+}\varepsilon,\\
 {\cal M}_{vu}\varepsilon &=& {\cal B}_{j}\gamma^{j+}\varepsilon,\\
 {\cal M}_{ui}\varepsilon &=&\left({\cal C}_{ij}\gamma^{j}P_-+{\cal E}\gamma^++{\cal G}_{ij}\gamma^{j+}\right)\varepsilon,\\
 {\cal M}_{ij}\varepsilon &=& \left({\cal I}_{ijk}\gamma^{k}P_-+{\cal
J}\gamma^++{\cal K}_{ijk}\gamma^{k+}\right)\varepsilon.
\end{eqnarray}
The expressions for quantities ${\cal A}$, ${\cal B}$, ${\cal C}$,
${\cal E}$, ${\cal G}$, ${\cal I}$, ${\cal J}$ and ${\cal K}$ are
quite lengthy and unimportant for our problem, so we do not
present them here. They involve functions $H(u,{\bf x})$, ${\bf
a}(u,{\bf x})$ and ${\cal F}(u,{\bf x})$ and their derivatives. In
particular, when the field strength vanishes, only ${\cal B}$ and
${\cal G}$ are nonzero and thus the holonomy group is the familiar $\R^3$
appropriate to the pseudo-Riemannian background.

In more general backgrounds the combinations of
the gamma matrices involved in the commutators are $\gamma^iP_-$,
$\gamma^+$ and $\gamma^{i+}$. Using this fact we define the
complete set of the holonomy generators of the null solution as
follows
\begin{eqnarray}
T^i&=&-\frac i2 P_-\gamma^iP_-, \\
R^i&=&P_+\gamma^{i+}P_-,\\
R^4&=&P_+\gamma^+P_-.
\end{eqnarray}
The generators $T^i$ generate an $SU(2)$ algebra because
$[T^i,T^j]=i\epsilon_{ijk}T^k$. Since $(\gamma^+)^2=0$ the
generators $R^i$ and $R^4$ commute with each other forming $\R^4$.
Choosing $T^3$ as the Cartan generator for $SU(2)$, we find that
the set of generators $\{R^i,R^4\}$ has weights $\pm\frac 12$ and
thus the pairs $\{R^1,R^2\}$ and $\{R^3,R^4\}$ transform as two
doublets under the action of $SU(2)$. This leads us to the
conclusion that the generalized holonomy group of the null
solution is
\begin{equation}
{\cal H}_{null}=SU(2)\ltimes 2\R^{2}.
\end{equation}

\section{Discussion}
\label{discuss}
\subsection{Relation to the $G$-structure}
As we have seen in the previous section, the generalized holonomies of
both timelike and null solutions preserving half of the supersymmetries
in $D=5$ are the same, namely $SU(2)\ltimes2{\R}^2$.  As shown in
\cite{Gauntlett}, the corresponding $G$-structures are $SU(2)$ for
timelike solutions and $R^3$ for null ones.  Both are subgroups of the
generalized holonomy group.  This result may be expected for the reason that
the $G$-structure is a global reduction of the frame bundle with
structure group $Spin(1,4)$ over five-dimensional spacetime to a
sub-bundle with structure group $G$ over a base manifold $\cal B$.

In section \ref{gh8susy} we have shown that the generalized holonomy group
for $n=4$ must be contained in a subgroup $SL(1,\H)\ltimes {\H}$ of the
$generalized$ structure group $SL(2,\H)$.  Recall that $SL(1,\H)\simeq SU(2)$
\cite{Gilmore}, and furthermore that $\H$ in the semi-direct product can
in fact be seen as two doublets $2{\R}^2$ under $SU(2)$.
To see this, recall that a quaternion $Q=q^0+iq^1+jq^2+kq^3\in
\H$, where $q^i\in\R$, can be written as
\begin{equation}
W=\pmatrix{q^0+iq^3 & q^1+iq^2 \cr -q^1+iq^2 & q^0-iq^3},
\end{equation}
where $W\in U(2)$ and $det(W)=\|Q\|$.  The action of $SL(1,\H)$ on $\H$ in
the semi-direct product is by left multiplication on $W$, and hence the
columns necessarily transform as doublets under $SU(2)$.


%
%
%

This demonstrates that the generalized holonomy for solutions preserving
$n=4$ supersymmetries is in fact $SL(1,\H)\ltimes{\H}\subset
SL(2,\H)$, where $\H$ becomes two doublets if the isomorphic group $SL(1,\H)
\simeq SU(2)$ is concerned.  Thus the solutions of $D=5$ minimal supergravity
support the classification scheme of table~\ref{gh}.  In addition, the
$G$-structure is embedded in the manner
$SU(2)\subset SU(2)\ltimes{2\R}^{2}\simeq SL(1,\H)\ltimes\H \subset SL(2,\H)$
for timelike solutions and
${\R}^3\subset\R^4\subset SU(2)\ltimes2{\R}^{2}\simeq SL(1,\H)\ltimes\H
\subset SL(2,\H)$
for null solutions.  We also notice that $SL(2,\H)\simeq Spin(1,5)$,
which is the structure group of six-dimensional spacetime.  That
is why $G$-structure only finds a unified picture in six dimensions
but not in five dimensions \cite{Gauntlett:01}.

\subsection{More on the generalized holonomy}
In the previous section we have seen how the same generalized
holonomy group $SU(2)\ltimes 2\R^2$ arises for two different
classes of supergravity vacua in five dimensions. It is
interesting to see how this works for other theories with 8
supercharges, i.e. $D=6$, $\mathcal N=(1,0)$ and $D=4$, $\mathcal N=2$
\cite{Gauntlett:01,Tod}. All three of them are expected to share a
common framework from either the $G$-structure or generalized holonomy
points of view. Indeed it has been found that six-dimensional
minimal supergravity has only null solutions with $G$-structure
$SU(2)\ltimes {\R}^{4}$, which contains the $G$-structures of the four and
five dimensional cases as subgroups \cite{Gauntlett:01}.

As for the generalized holonomy, for timelike solutions of the
four dimensional theory, from the integrability conditions we
obtain the complete set of holonomy generators
$\{\gamma^{ij}P_-,\gamma^iP_-,\gamma^5P_-\}$, where $i,j=1,2,3$
and $\gamma^5=\gamma^{0123}, P_-=\frac{1}{2}(1-\gamma^0)$. In the
null case, we have generators
$\{\gamma^{i+}P_-,\gamma^+P_-,\gamma^+\gamma^5P_-,\gamma^iP_-,\gamma^5P_-\}$,
where $i=2,3$ and $\gamma^5=\gamma^{+-23}$,
$P_-=\frac{1}{2}\gamma^-\gamma^+$. For $D=6$, $\mathcal N=(1,0)$,
we have generators $\{\gamma^{i+}P_-,\gamma^{ij}P_-\}$, where
$i=2,3,4,5$ and $P_-=\frac{1}{2}\gamma^-\gamma^+$ for only null
solutions. Hence for all cases we have the same generalized
holonomy group $SU(2)\ltimes 2\R^2\simeq SL(1,\H)\ltimes \H$ as
was expected in section \ref{gh8susy}. Our results are summarized
in table \ref{t} together with the $G$-structures found in
\cite{Gauntlett,Gauntlett:01}. Depending on the particular
solution, the generalized holonomy may be a subgroup of
$SU(2)\ltimes 2\R^2$. For example, for solutions with vanishing
flux the holonomy group is restricted to the G-structure group
which is indeed a subgroup of $SU(2)\ltimes 2\R^2$.

It would be interesting to further test this conjecture on other vacua
also with $8$ supercharges, such as $D=4,5$ supergravity coupled
with matter multiplets. Since they can be obtained from $D=6$,
$\mathcal N=(1,0)$ (without truncation), one may expect their generalized
holonomy would be the same as what we found here. Another
interesting test could be done with the gauged supergravity vacua \cite{Klemm,Gauntlett:02}.
In this case, as we have seen in section \ref{gh8susy}, the
structure group is $SL(4,\C)$ and thus it is possible to stabilize
0, 1, 2, 3 or all 4 complex-valued spinors, depending on the
solution. This means that a given solution may preserve $n=0,2,4,6$ or $8$
(full) supersymmetry. However, we are reminded that
in $D=11$ supergravity, although all values of preserved
supersymmetries from $0$ to $32$ are all allowed by the M-algebra, not
all of them are found \cite{Duff:01,Hull,Townsend}. The fact that
solutions to gauged $D=4,5$ supergravity preserving $n=6$ are not
found \cite{Klemm,Gauntlett:02} may be similar to the conjectured absence of a solution
with 31 supersymmetries in conventional $D=11$ supergravity theory, although
an argument for their existence can be made \cite{preons}.

\begin{table}
\begin{center}
\begin{tabular}{cccccc}
Dim&Solution&Gen. Structure&
Generalized &$G$-structure\vspace{-8pt}\\ &Type&Group&
Holonomy&\\
\hline
4&timelike&$SL(2,\H)$&
$SU(2)\ltimes 2{\R}^{2}$&$SU(2)$\\
4&null&$SL(2,\H)$&
$SU(2)\ltimes 2{\R}^{2}$&$\R^2$\\
5&timelike&$SL(2,\H)$&
$SU(2)\ltimes 2{\R}^{2}$&$SU(2)$\\
5&null&$SL(2,\H)$&
$SU(2)\ltimes 2{\R}^{2}$&$\R^3$\\
6&null&$SL(2,\H)$&
$SU(2)\ltimes 2{\R}^{2}$&$SU(2)\ltimes{\R}^4$
\end{tabular}
\end{center}
\caption{Generalized holonomy groups for half-BPS vacua of supergravity
with $8$ supercharges and the corresponding $G$-structures.} \label{t}
\end{table}

\section*{Acknowledgments}
We are grateful to James T. Liu and Michael Duff for most useful
discussions.

\bibliographystyle{amsplain}

\end{document}